\documentclass[reprint,twocolumn,pra,nofootinbib]{revtex4-1}
\usepackage{amsmath,amssymb}
\usepackage{epsfig}
\usepackage[utf8]{inputenc}
\usepackage{amsmath,amsfonts}
\usepackage{graphicx}
\usepackage{color}
\usepackage{hyperref}
\usepackage{turnstile}
\usepackage{relsize}
\usepackage[toc,page]{appendix}
\usepackage{pstricks-add}
\usepackage{relsize}
\usepackage{mathtools}
\usepackage{physics}
\usepackage{dcolumn}
\newcommand{\ru}[3]{\rule[#1mm]{#2mm}{#3mm}}
\begin{document}
\title{Unravelling Interaction and Temperature Contributions\\ in Unpolarized Trapped Fermionic Atoms in the BCS Regime}
\author{Sejung Yong}\thanks{Both authors contributed equally.}
\affiliation{Physics Department and Research Center OPTIMAS, Rhineland-Palatinate Technical University Kaiserslautern-Landau,
Erwin-Schr\"odinger Straße 46, 67663 Kaiserslautern, Germany\\
{\tt yong@rptu.de, cbarbosa@physik.uni-kl.de, jekoch@rptu.de, langf@rptu.de, axel.pelster@rptu.de, widera@physik.uni-kl.de}}
\author{Sian Barbosa}\thanks{Both authors contributed equally.}
\affiliation{Physics Department and Research Center OPTIMAS, Rhineland-Palatinate Technical University Kaiserslautern-Landau,
Erwin-Schr\"odinger Straße 46, 67663 Kaiserslautern, Germany\\
{\tt yong@rptu.de, cbarbosa@physik.uni-kl.de, jekoch@rptu.de, langf@rptu.de, axel.pelster@rptu.de, widera@physik.uni-kl.de}}
\author{Jennifer Koch}
\affiliation{Physics Department and Research Center OPTIMAS, Rhineland-Palatinate Technical University Kaiserslautern-Landau,
Erwin-Schr\"odinger Straße 46, 67663 Kaiserslautern, Germany\\
{\tt yong@rptu.de, cbarbosa@physik.uni-kl.de, jekoch@rptu.de, langf@rptu.de, axel.pelster@rptu.de, widera@physik.uni-kl.de}}
\author{Felix Lang}
\affiliation{Physics Department and Research Center OPTIMAS, Rhineland-Palatinate Technical University Kaiserslautern-Landau,
Erwin-Schr\"odinger Straße 46, 67663 Kaiserslautern, Germany\\
{\tt yong@rptu.de, cbarbosa@physik.uni-kl.de, jekoch@rptu.de, langf@rptu.de, axel.pelster@rptu.de, widera@physik.uni-kl.de}}
\author{Axel Pelster}
\affiliation{Physics Department and Research Center OPTIMAS, Rhineland-Palatinate Technical University Kaiserslautern-Landau,
Erwin-Schr\"odinger Straße 46, 67663 Kaiserslautern, Germany\\
{\tt yong@rptu.de, cbarbosa@physik.uni-kl.de, jekoch@rptu.de, langf@rptu.de, axel.pelster@rptu.de, widera@physik.uni-kl.de}}
\author{Artur Widera}
\affiliation{Physics Department and Research Center OPTIMAS, Rhineland-Palatinate Technical University Kaiserslautern-Landau,
Erwin-Schr\"odinger Straße 46, 67663 Kaiserslautern, Germany\\
{\tt yong@rptu.de, cbarbosa@physik.uni-kl.de, jekoch@rptu.de, langf@rptu.de, axel.pelster@rptu.de, widera@physik.uni-kl.de}}
\begin{abstract}
In the BCS limit density profiles for unpolarized trapped fermionic clouds of atoms are largely featureless. Therefore, it is a delicate task to analyze them in order to quantify their respective interaction and temperature contributions. Temperature measurements have so far been mostly considered in an indirect way, where one sweeps isentropically from the BCS to the BEC limit. Instead we suggest here a direct thermometry, which relies on measuring the column density and comparing the obtained data with a Hartree-Bogoliubov mean-field theory combined with a local density approximation. In case of an attractive interaction between two-components of $^{6}$Li atoms trapped in a tri-axial harmonic confinement we show that minimizing the error within such an experiment-theory collaboration turns out to be a reasonable criterion for analyzing in detail measured densities and, thus, for ultimately determining the sample temperatures. The findings are discussed in view of various possible sources of errors.
\end{abstract}
\maketitle
{\it Introduction.}\,--
In equilibrium strongly correlated fermions can be in a normal or a superfluid state depending on the temperature and the two-particle interaction strength. The phase diagram contains a crossover between the limiting cases of Bose-Einstein condensates of molecules (BEC) and Cooper pairs of fermions (BCS). The critical line in the phase diagram between normal and superfluid was theoretically predicted both for a homogeneous and for a harmonically trapped unpolarized  Fermi gas in 1993 \cite{Melo1993} and in 2004 \cite{Perali2004}, respectively. But until now it has remained to be experimentally elusive to directly measure the critical temperature for a given interaction strength \cite{Ketterle2008,Onofrio2016}.

Most delicate is, in particular, the BCS limit as then the density profiles are quite featureless. Thus, it becomes difficult to reconstruct from measured density profiles the respective temperature and interaction contributions.
For a long time the most promising temperature measurement in the BCS limit relies on an indirect way in form of an adiabatic sweep thermometry \cite{Regal2004,Bartenstein2004,Carr2004,Chen2005}. The underlying idea is to change the s-wave scattering length via a Feshbach resonance with an isentropic sweep from the BCS to the BEC regime. Measuring the shape of the profile of the trapped cloud deep in the BEC regime allows to determine the temperature by taking a Hartree-Fock approach into account \cite{Giorini1996}. 
A theoretically obtained entropy-temperature gauge curve then allows to infer from the entropy known in the BEC regime the original temperature in the BCS regime. Furthermore, by performing a control sweep back from the BEC to the BCS limit one can check experimentally whether the cloud properties are changed reversibly. 
Alternatively, exploiting the universal thermodynamics around unitarity~\cite{Horikoshi10, Navon10, Ku12}, the temperature can be inferred. In particular, the high-momentum tails of the density distribution after time of flight reveal the so-called contact~\cite{Tan08} and can be used to determine the temperature~\cite{Kuhnle11}.

Recently the homogeneous phase diagram of the BEC-BCS crossover was precisely mapped out with a method, which is based on an artificial neural network \cite{Koehl2023,Koehl2023b}. By applying advanced image recognition techniques to the momentum distribution of the fermions, which has been widely considered as featureless for providing information about the condensed state, the critical temperature was measured. With this the long-standing prediction of a maximum on the bosonic side of the crossover \cite{Melo1993} was confirmed in accordance with the extended Gorkov-Melik-Bakhudarov theory \cite{Pisani2018}.

In this experiment-theory collaboration we work out a direct thermometry for harmonically trapped fermionic atoms in the BCS limit. 
It is based on extending the usual Bogoliubov mean-field theory by an additional contribution from the Hartree channel, which physically corresponds to the interaction energy of all Cooper pairs, see e.g.~\cite{Urban,Koehl2015}. 
The full profile of a measured \textit{in-situ} absorption image is then compared with theoretically determined densities for different temperatures. The resulting mean-squared error turns out to have a unique minimum in the BCS limit \(1/(k_{\rm F} a_{\rm s})< -1\), which allows to determine the temperature of the sample. In addition, we can discriminate between the normal and the superfluid phase by having either a simultaneous minimum for both the Hartree- and the Hartree-Bogoliubov theory in a normal fluid or only for the Hartree-Bogoliubov theory as is the case for a superfluid. 
Furthermore, towards unitarity, i.e. for \(1/(k_{\rm F} a_{\rm s}) \ge -1\), this direct thermometry leads to inconsistencies indicating its breakdown, as one enters a regime where a mean-field theory is no longer valid and would have to be extended by fluctuation corrections. 

{\it Experiment.\,--}
We create degenerate quantum gases of $^6$Li atoms prepared in equal amounts in the two lowest-lying Zeeman substates of the eletronic ground state $^2S_{1/2}$ using standard techniques of laser and evaporative cooling, see Refs.~\cite{Gaenger_apparatus, Nagler_cloudshape} for further experimental details. Evaporation takes place inside a hybrid magnetic-optical trap in the vicinity of the broad Feshbach resonance centered at \SI{832.2}{G}, which enables us to tune the $s$-wave scattering length~\cite{Grimm_review, Zuern_resonance}. 
With the laser power used to create the optical dipole trap (ODT) and reached at the end of the evaporation scheme, we control the sample's temperature $T$ and trap frequencies $\omega_i$ along the three axes $x_i$ with $i = 1, 2, 3$. More specifically, reducing the final ODT power decreases these quantities as well as the atom number simultaneously.
\begin{figure}[t]
\includegraphics[width=86mm]{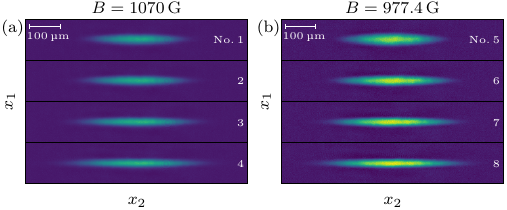}
\caption{
    Column-integrated absorption images. Shown are the four images for both magnetic fields analyzed in this work, see Tab.~\ref{tab:data} for an overview of the respective settings. Color scale has been normalized to the maximum and minimum recorded density of all images. From top to bottom: (a)~images 1 to 4, $B = \SI{1070}{G}$, (b)~images 5 to 8, $B = \SI{977.4}{G}$. 
} \label{fig:images}
\end{figure}
We typically achieve temperatures of the order of $T \approx \SI{100}{\nano K}$ on the bosonic side of the resonance, i.e. $1 / (k_{\rm F} a_{\rm s}) > 0$, which is measured by fitting a bimodal density distribution to the image data deep in the BCS regime, at a magnetic field around $B = \SI{680}{G}$~\cite{Nagler_cloudshape, Nagler_dynamicdisorder}. 
In this work, we investigate samples of up to $N = 10^6$ atoms in harmonic traps with frequencies between $(\omega_1, \omega_2, \omega_3) = 2\pi\cdot(210, 25.5, 129)\,\text{Hz}$ and $2\pi\cdot(386, 26.6, 248)\,\text{Hz}$. 
Note that $\omega_2$ effectively depends only on $B$~\cite{Gaenger_apparatus, Nagler_cloudshape} and does not change significantly over the range investigated here. 
We present two data sets, which have been taken at two different magnetic fields for respectively four final trap configurations, one at $B = \SI{1070}{G}$ and the other at $B = \SI{977.4}{G}$. At the end of every sequence run, the atom sample is imaged \textit{in-situ} by performing high-intensity absorption imaging~\cite{Reinaudi_imaging} along the $x_3$-direction after allowing it to thermalize for \SI{100}{\milli\second}. In the series measured at \SI{977.4}{G}, we used ten repetitions for each trap configuration, while the series at \SI{1070}{G} used 67 repetitions per setting. We average all images taken for the same setting and compute the atom number by integrating over the entire averaged absorption image. For the uncertainty of the atom number, we use the standard deviation of the atom number as calculated from individual images, which is typically of the order of \SI{5}{\%} of $N$. The averaged absorption images can be seen in Fig.~\ref{fig:images}, whereas Tab.~\ref{tab:data} collects the respective system parameters like the atom numbers, the trap frequencies, and the interaction strengths for the column densities. Finally, we infer the line density $n_{\rm ex}(x_2)$ by integrating the averaged absorption image along the $x_1$-axis. A detailed theoretical analysis shows that they can not be described by the usual BCS mean-field theory alone. In order to use these measured column densities for a reliable thermometry it turns out to be mandatory to consider instead all mean-field interaction channels. In the rest of this letter we show that this yields the sample temperatures in the last column of Tab.~\ref{tab:data}.
\begin{table}[t]
\resizebox{\columnwidth}{!}{%
    \begin{tabular}{c|c|c|c|c|c}
        No. & \(B/\)\text{G} & $(\omega_1, \omega_2, \omega_3) \, / \, (2 \pi \text{Hz})$ & $N/10^5$ & $1/(k_\mathrm{F} a_\mathrm{s})$ & $T/T_\mathrm{F}$\\
        \hline
       \ru{0}{0}{4}1 & 1070 & (243, 26.6, 152) & $4.8 \pm 0.2$ & $-1.505 \pm 0.011$ & $0.047 \pm 0.002$ \\[0.5mm]
        \ru{0}{0}{3}2 & 1070 & (272, 26.6, 171) & $5.2 \pm 0.2$ & $-1.428 \pm 0.010$ & $0.103 \pm 0.001$ \\[0.5mm]
         \ru{0}{0}{3}3 & 1070 & (323, 26.6, 205) & $5.8 \pm 0.2$ & $-1.324 \pm 0.008$ & $0.187 \pm 0.001$ \\[0.5mm]
       \ru{0}{0}{3}4 & 1070 & (386, 26.6, 248) & $6.4 \pm 0.2$ & $-1.226 \pm 0.007$ & $0.257 \pm 0.001$ \\[0.5mm]
     \ru{0}{0}{3}5 & 977.4 & (210, 25.5, 129) & $7.9 \pm 0.4$ & $-1.119 \pm 0.010$ & $0.073 \pm 0.010$ \\[0.5mm]
     \ru{0}{0}{3}6 & 977.4 & (272, 25.5, 171) & $8.4 \pm 0.5$ & $-1.010 \pm 0.009$ & $0.115 \pm 0.001$ \\[0.5mm]
       \ru{0}{0}{3}7 & 977.4 & (323, 25.5, 205) & $8.7 \pm 0.6$ & $-0.949 \pm 0.010$ & $0.182 \pm 0.002$ \\[0.5mm]
      \ru{0}{0}{3}8 & 977.4 & (386, 25.5, 248) & $9.0 \pm 0.6$ & $-0.887 \pm 0.009$ & $0.245 \pm 0.002$ \\
    \end{tabular}
    }
    \caption{Experimental data for the eight measured absorption images shown in Fig.~\ref{fig:images}.}
    \label{tab:data}
\end{table}

{\it Mean-Field Equations.}\,--
Trapped fermions with mass $M$ and two spin species \(\sigma =\uparrow, \downarrow\)  are described in second quantization by the Hamilton operator
\begin{eqnarray}
\label{Hamiltonian}
	\hat H
	&=&
	\int d^3 x
	\Bigg\{
	\sum_{\sigma}
	\hat \psi_\sigma^\dagger(\vb* x)
	\left[
	- \frac{\hbar^2}{2 M}\, \Delta
	- \mu
    + V_{\rm HO}(\vb* x)
	\right]
	\hat \psi_{\sigma}(\vb* x)  \nonumber\\
&&	+
	g
	\hat \psi_\uparrow^\dagger(\vb* x)
	\hat \psi_\downarrow^\dagger(\vb* x)
	\hat \psi_\downarrow(\vb* x)
	\hat \psi_\uparrow(\vb* x)
	\Bigg\}\, .
\end{eqnarray}
Here \(V_{\rm HO}(\vb* x)=(M/2)\sum_{i=1}^3 (\omega_i x_i)^2\) denotes a harmonic trapping potential, \(\mu\) is the chemical potential, and \(g\) stands for the contact interaction strength, whereas $\hat \psi_\sigma^\dagger(\vb* x)$ and $\hat \psi_{\sigma}(\vb* x)$ represent creation and annihilation field operators, respectively, obeying the fermionic anti-commuator algebra. As the Hamilton operator (\ref{Hamiltonian}) can not be diagonalized, a mean-field approach is needed. The most general mean-field theory is provided by the Hartree-Fock–Bogoliubov (HFB) theory, which takes unbiasedly all three possible mean-field channels into account. However, the Supplemental Material~\cite{SM} shows that the Fock channel of the contact interaction does not contribute due to its vanishing effective range \cite{Czycholl2023}. Therefore, the HFB theory reduces to a Hartree-Bogoliubov (HB) mean-field theory. Furthermore, we only consider the unpolarized case, i.e.~we assume the same number of fermionic atoms for both spin species \(\sigma = \uparrow, \downarrow\). 
For this case the Supplemental Material derives the corresponding HB mean-field equations \cite{SM}, which we just list here. The given particle number $N= \int d^3 x \,n(\vb* x)$
fixes the chemical potential $\mu$ with the density being determined self-consistently from
\begin{equation}\label{eq: density eq}
    n(\vb* x)
    =
   \int \frac{d^3 k}{(2 \pi)^3}
    \left[
    1
    -
    \frac{\epsilon_k -\mu_{\rm H}(\vb* x)}{E_k(\vb*x)}
    \tanh{\frac{\beta E_k(\vb*x)}{2}}
    \right]\, .
\end{equation}
Here the local HB dispersion is given by
$
E_k(\vb* x)
=
\sqrt{\left[ \epsilon_k -\mu_{\rm H} (\vb* x) \right]^2 +\Delta^2 (\vb* x)}
$
with the local chemical potential
$\mu_{\rm H}(\vb* x)
    =
    \mu
    -\Gamma^{\rm H} (\vb* x)
    -
    V_{\rm HO}(\vb* x)$
and the local Hartree energy 
$\Gamma^{\rm H} (\vb* x)= g n(\vb* x)/2$.
Correspondingly, the local Bogoliubov energy $\Delta(\vb* x)$ follows from
\begin{equation}\label{eq: gap eq renormalized}
    \frac{M}{2\pi \hbar^2 a_{\rm s}}
    =
   \int \frac{d^3 k}{(2 \pi)^3}
    \left[
    \frac{1}{\epsilon_k}
    -
    \frac{1}{E_k(\vb*x)}
    \tanh{\frac{\beta E_k(\vb*x)}{2}}
    \right]
    \, ,
\end{equation}
where $a_{\rm s}$ denotes the s-wave scattering length between spin-up and -down fermions. Solving the above set of mean-field equations allows to determine the density profile for a given set of experimental parameters.

{\it Data Analysis.}\,--
We analyse the \textit{in-situ} absorption images shown in Fig.~\ref{fig:images} and tabled in Tab.~\ref{tab:data} based on the column density profile \(n_{\rm ex}(x_2)\). Taking into account the experimental parameters, we determine the squared error between the measured column density and the density profile $n(x_2)$ calculated from the above HB mean-field equations according to
\begin{equation}\label{eq: squared error}
    \Delta n(x_2)
    =   \sqrt{\sum_{m=1}^{N_2}\frac{\left[n^{(m)}(x_2) -n_{\rm ex}^{(m)} (x_2)\right]^2}{N_1}}\, .
\end{equation}
Here \(m\) indexes a measured point and \(N_i\) denotes the number of measured points in direction $i$. Apart from the experimentally accessible parameters like the inverse dimensionless interaction strength \(1/(k_{\rm F} a_{\rm s})\), the particle number $N$, and the trap frequencies \(\omega_1,\omega_2,\omega_3\), there is additionally one unknown parameter, namely the temperature \(T\). Therefore, the error 
(\ref{eq: squared error}) is calculated with the theoretical column density \(n(T, x_2)\) for varying temperatures \(T\). If there exists a unique minimum of the temperature-squared error curve, we can read off from it the temperature of the trapped Fermi gas sample. Thus, we suggest here a direct thermometry on the basis of combining experimental and theoretical methods.
\begin{figure}[t]
\includegraphics [width=0.5\textwidth] {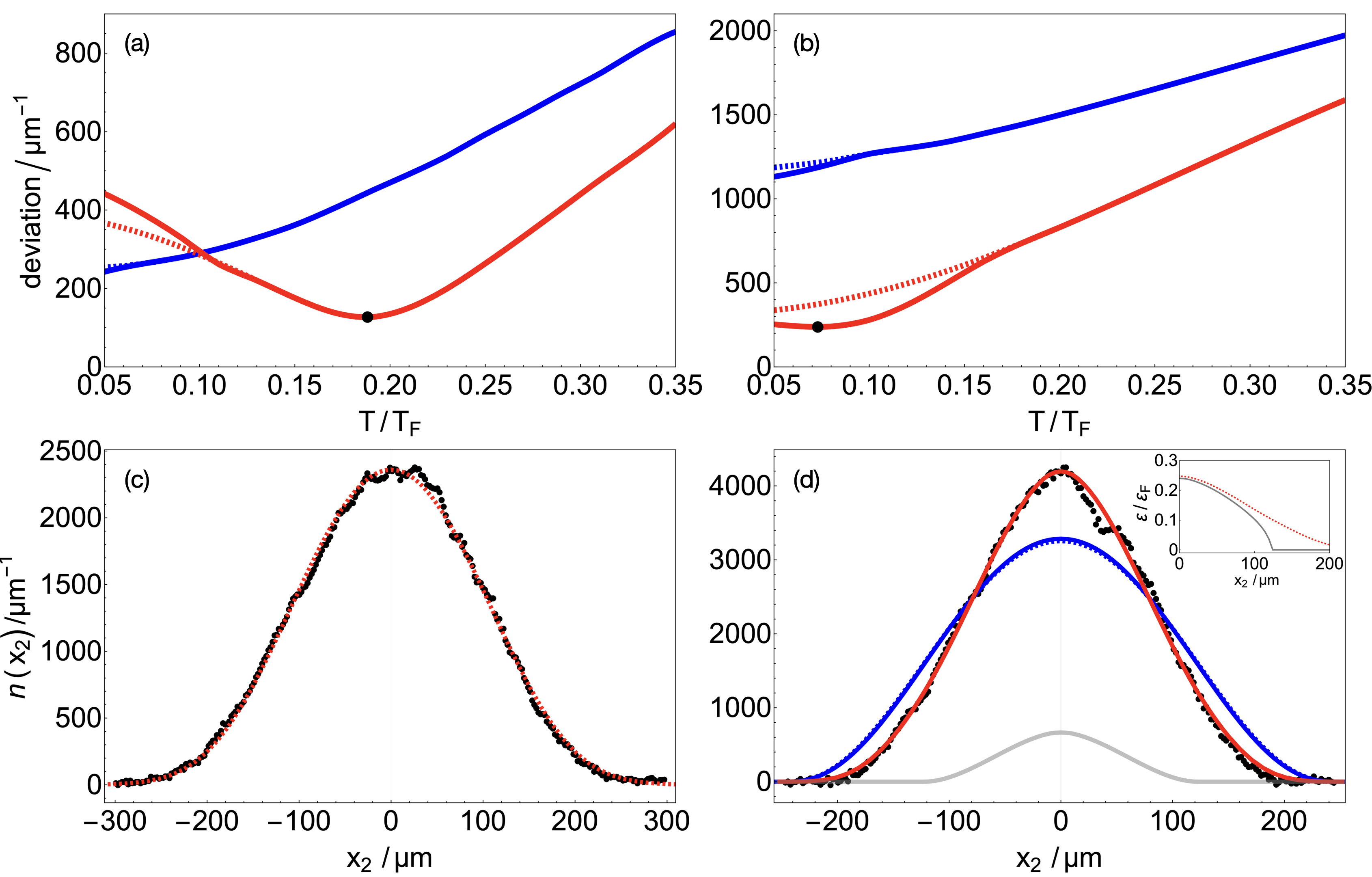} \centering
\caption{
Thermometry with unique error minimum.
(a–b) Curves of squared error between measured (dots) and theoretical column densities on \(x_2\)-axis at various temperatures based on ideal Fermi gas model (dashed blue), Bogoliubov theory (blue), Hartree theory (dashed red), and HB theory (red).
(a) Sample 3 has a mutual error minimum from the Hartree- and the HB theory.
(b) Sample 5 has an error minimum from the HB theory.
(c) Fitted column density profile of normal fluid sample 3 from experiment by Hartree- as well as HB theory.
(d) Fitted column density profile of superfluid sample 5 from experiment (dots) and HB theory as well as condensate column density (gray) from (\ref{CD}). For comparison also density profiles from ideal Fermi gas and Bogoliubov theory at $T=0$ are shown.
Inset: Absolute values of Hartree energy $\langle \Gamma^{\rm H} (x_2) \rangle$ (dashed red) and gap energy $\langle \Delta (x_2) \rangle$ (gray) averaged in $x_1$-$x_3$ plane.}\label{fig: thermoemtry samples}
\end{figure}

We illustrate this thermometry at first by the example of Fig.~\ref{fig: thermoemtry samples}, where unique error minima occur. The sample 3 (Fig.~\ref{fig: thermoemtry samples}(a)) is identified to be normal fluid as both Hartree and HB theory turn out to have the same error minimum. And we read off from the location of the minimal squared error that the  temperature of the sample is $T = 0.187\,T_{\rm F}$. By contrast, sample 5 (Fig.~\ref{fig: thermoemtry samples}(b)) must be superfluid, since the Hartree theory does not have a minimum, but only the HB theory results in an error minimum. Furthermore, we read off that the minimal squared error occurs at the temperature $T=0.073\,T_{\rm F}$. Both for the normal and the superfluid sample we find for the respective temperature at the error minimum in Fig.~\ref{fig: thermoemtry samples}(c-d) a theoretical column density curve which perfectly lies on top of the experimental data points. Thus, we conclude that the proposed thermometry leads to a striking agreement. The existence of such an error minimum from Hartree- or HB theory is remarkable, since both ideal Fermi gas and BCS theory are incapable of fitting the experimental data even at zero temperature as is shown in Fig.~\ref{fig: thermoemtry samples}(d). For the BCS theory it is known that the interaction is only indirectly taken into account, thus after the line-of-sight integration the interaction effect is barely visible in the column density \cite{Zwierlein2006}. In order to take interaction effects into account in view of analysing experimental data reliably, it is thus mandatory to include the Hartree interaction term.

Once the temperature of a sample is determined, the theoretical modelling allows to extract further physically relevant quantities, which are not directly experimentally accessible. For the sake of comparison Fig.~\ref{fig: thermoemtry samples}(d) shows also the condensate column density \cite{Salasnich2005, Salasnich2013}:
\begin{eqnarray}
\label{CD}
n_0(x_2) = \int dx_1 dx_3 \Delta^2(\vb* x) \int \frac{d^3 k}{(2\pi)^3}   \frac{\tanh^2 [\beta E_k(\vb* x)/2]}{2 E_k^2(\vb* x)}
,
\end{eqnarray}
which yields the condensate fraction $N_0/N=0.102$ with $N_0=\int dx_2 n_0(x_2)$. And even more quantitative insight is obtained by considering Hartree and gap energies $\langle \Gamma^{\rm H}(x_2)\rangle$ and $\langle \Delta (x_2)\rangle$ averaged in the $x_1$-$x_3$ plane, see inset of Fig.~\ref{fig: thermoemtry samples}(d). 
The radius in $x_2$-direction of the area-averaged Hartree energy $r_{\rm H}$=218.2\,µm is approximately twice as large as the radius of the area-averaged gap energy $r_{\Delta}$=122.2\,µm. And we obtain that the area-averaged energies  have at $x_2=0$ the same absolute values $\langle \Delta(x_2 = 0) \rangle = 0.240 \,\epsilon_{\rm F}$ and $\langle \Gamma^{\rm H}(x_2=0) \rangle = -0.247\,\epsilon_{\rm F}$.
\begin{figure}[t]
\includegraphics [width=0.5\textwidth] {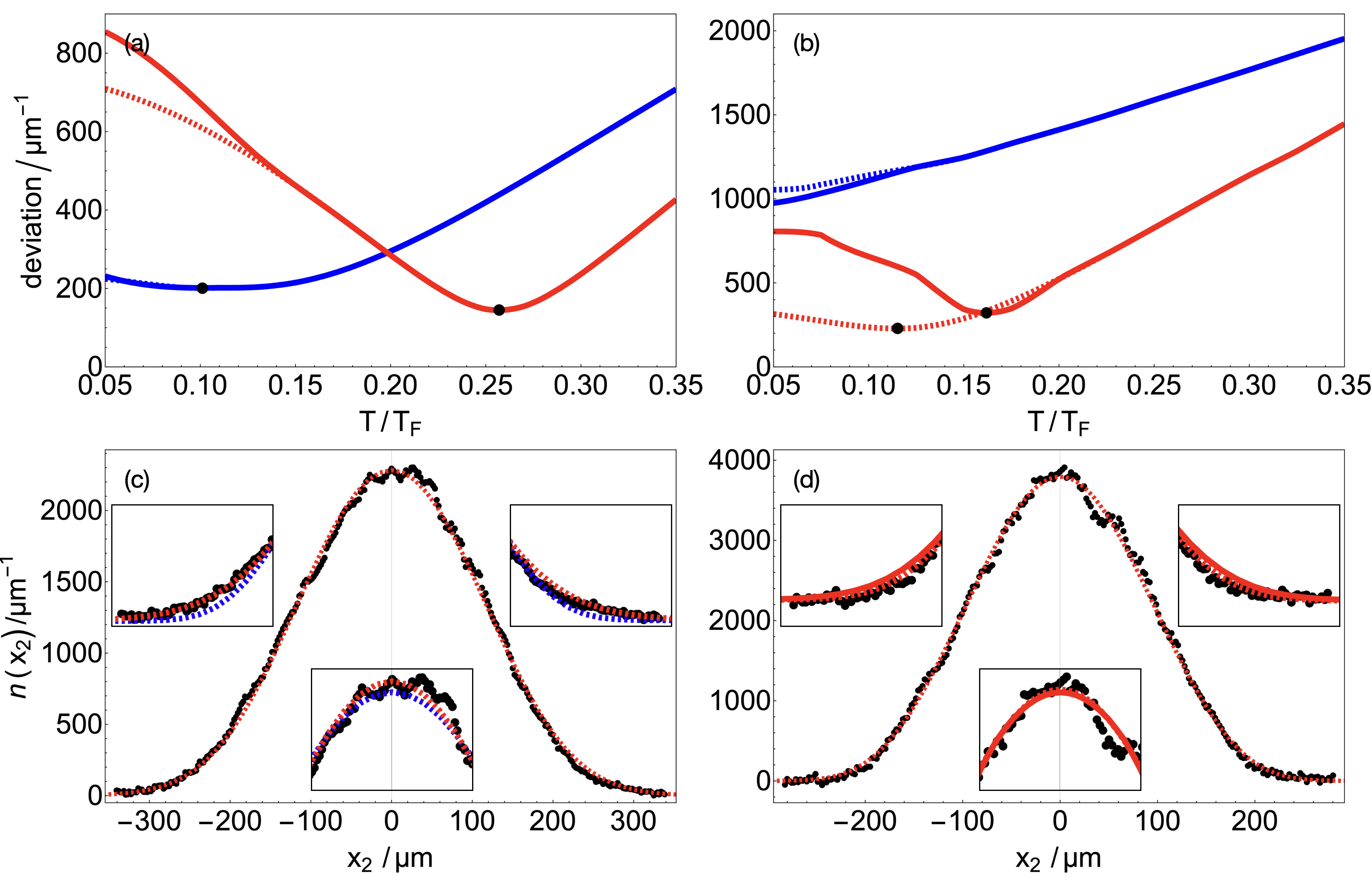}
\centering
\caption{Thermometry with non-unique error minima. (a–b) Curves of squared error between measured and theoretical column densities on \(x_2\)-axis at various temperatures based on ideal Fermi gas model (dashed blue), Bogoliubov theory (blue), Hartree theory (dashed red), and HB theory (red). (a) Sample 4 shows error minima from both ideal Fermi gas and Hartree theory. (b) Sample 6 has error minima from both Hartree and HB theory. 
(c) and (d) Fitted column density profile of normal fluid sample 4 and 6, respectively, from experiment (dots) and Hartree theory. Insets enlarge trap center and wings for comparing measured column density with theoretical curves evaluated at temperatures with minimal error.}
\label{fig: thermoemtry samples extra}
\end{figure}

But the data analysis is not always as straight-forward as discussed so far. Namely, it could happen that multiple error minima occur. In such a situation we identify the temperature of the sample with the one at the smallest error. For example, the sample 4 (Fig.~\ref{fig: thermoemtry samples extra}(a)) has squared error minima for the fit with both the ideal Fermi gas and the Hartree theory, which correspond to the temperatures $T = 0.101\,T_{\rm F}$ and $T = 0.257\,T_{\rm F}$, respectively. 
The latter temperature is the physically realized one due to the smaller error. 
Furthermore, the sample is identified to be a normal fluid due to the smallest error occuring for the Hartree theory. From the insets at both the trap center and the wings
we read off that the Hartree theory fits better than the ideal Fermi gas model in accordance with its smaller mean-squared error. 
Concerning the sample 6 (Fig.~\ref{fig: thermoemtry samples extra}(b)) we recognize again two error minima, this time due to the Hartree and the HB theory, yielding here the temperatures $T = 0.115\,T_{\rm F}$ and $T = 0.162\,T_{\rm F}$, respectively. The first value corresponds to the smaller error minimum and is, thus, to be considered as the result of our thermometry. Since the smallest error stems here from the Hartree theory, these data also correspond to the normal fluid. And, again the theoretical column densities in Fig.~\ref{fig: thermoemtry samples extra}(c-d) turn out to lie precisely on top of the experiment data points, documenting also in the case of non-unique error minima that our thermometry procedure works. From the respective insets we read off that
the theoretical densities evaluated at temperatures with minimal error differ more at the wings, whereas the differences are less visible at the center.
Note that  the small density modulations as well as the kink towards $+x_2$ seen in the experimental data of Figs.~\ref{fig: thermoemtry samples}(d) and Fig.~\ref{fig: thermoemtry samples extra}(d)
arise from fringes in the absorption image.
\begin{figure}[t]
\includegraphics [width=0.4\textwidth] {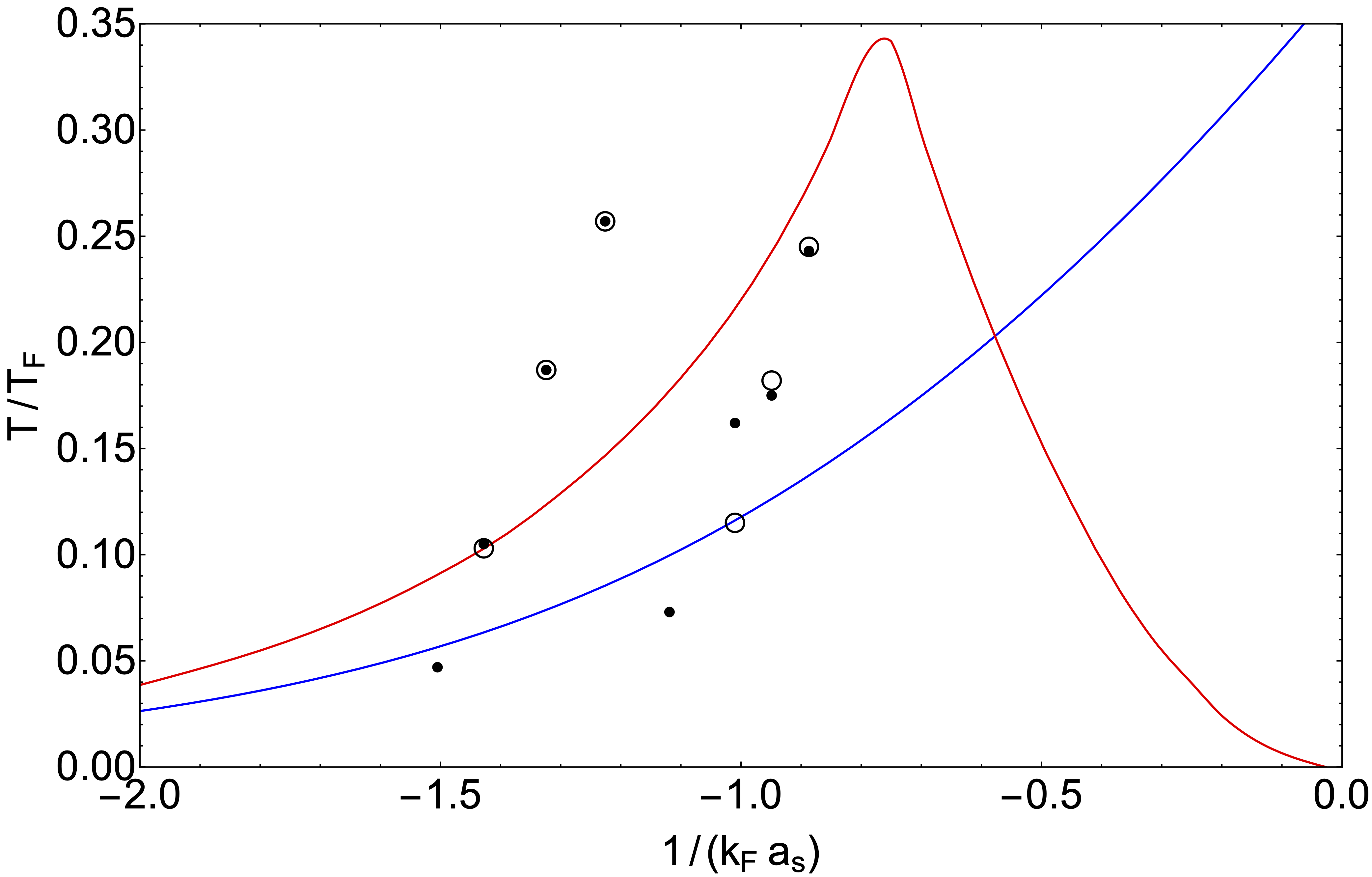} \centering
\caption{Thermometry by Hartree (circle) and HB (point) theory for harmonically trapped $^{6}$Li atoms at interaction strenghts corresponding to sample 1–8 of Tab.~\ref{tab:data} from left to right. Furthermore, critical line between normal fluid and superfluid according to Bogoliubov (blue curve) \cite{Perali2004} and HB (red curve) theory is shown.} \label{fig: phasediagram}
\end{figure}

In total we analysed the measured column densities for 8 samples and identified their respective temperatures on the basis of both the Hartree- and the HB theory as depicted in Fig.~\ref{fig: phasediagram}. Therein we also plot for harmonically trapped $^{6}$Li atoms the critical temperature \(T_{\rm c}\) between normal fluid and superfluid as a function of the inverse interaction strength \(1/(k_{\rm F} a_{\rm s})\). 
As the onset of superfluidity occurs at the trap center, \(T_{\rm c}\) follows from evaluating the condition \(\Delta(\vb* x=0)=0\) by taking into account the above mean-field equations. Neglecting the Hartree term reproduces the Bogoliubov transition temperature from Ref.~\cite{Perali2004}, otherwise we obtain the HB transition temperature. In the BCS limit (\(1/(k_{\rm F} a_{\rm s}) < -1\)) the critical temperature of the HB theory is higher than the one from the Bogoliubov theory due to a larger density at the trap center. Towards unitarity we observe that the critical temperature from the HB theory has a maximum, which occurs once the absolute value of the Hartree energy coincides with the Fermi energy. However, even before this maximal critical temperature the HB theory leads to an inconsistent thermometry in the strongly interacting regime \(1/(k_{\rm F} a_{\rm s})\approx -1\). This can read off from the experimental data analysed in  Fig.~\ref{fig: thermoemtry samples extra}(b),(d). There it turns out that the Hartree theory has the smallest error at a temperature, which lies  in the superfluid phase according to HB theory, see Fig.~\ref{fig: phasediagram}. This represents an inconsistent result as below the critical temperature the Bogoliubov  gap parameter should be non-vanishing. The same inconsistency also occurs for two other samples, which were measured for stronger interactions, see Fig.~\ref{fig: phasediagram}. We attribute these inconsistencies to the fact that our mean-field thermometry has a limited range of validity, which is restricted to the BCS limit. 
Thus, in order to make the thermometry even more reliable and to extend it towards or even beyond unitarity, it is therefore necessary to work out a corresponding beyond-mean-field description.

{\it Summary and Outlook.}\,--
Within an experiment-theory collaboration we were able to quantitatively identify the respective interaction and temperature contributions in the density profiles for unpolarized trapped interacting fermions in the BCS limit. This is insofar a striking result as the experimentally measured densities are largely featureless, see Fig.~1. In particular, we established a direct thermometry by identifying the temperature of a sample with the minimal error between experimentally measured column densities and corresponding ones calculated on the basis of the HB mean-field theory. The theoretical density profiles generically consist of an inner superfluid and an outer normal fluid part determined from a HB and a Hartree theory, respectively. This procedure is analogous to the widely accepted bimodal fit method for the condensed molecules at the BEC side \cite{Bartenstein2004}. Furthermore, the theoretically determined column densities turn out to lie precisely on top of the measured ones. Therefore, our findings shed new light on the importance of the Hartree contribution in the BCS limit, which is neglected in the usual BCS theory. And, we deduced from analyzing the experimental data that the HB theory fails towards unitarity due to its mean-field nature. In order to work out a thermometry over the whole BCS-BEC crossover range, beyond mean-field corrections would have to be taken into account.

{\it Acknowledgement.}\,-- We thank Andr\'e Becker, Nikolai Kaschewski, Corinna Kollath, and Carlos Sá de Melo for valuable discussions as well as acknowledge financial support by the Deutsche Forschungsgemeinschaft (DFG, German Research Foundation) via the Collaborative Research Center SFB/TR185 (Project No.~277625399). J.K. was supported by the Max Planck Graduate Center with the Johannes Gutenberg-Universität Mainz.

{\it Supplemental Material.}\,-- Here we work out a mean-field approach for trapped fermions. To this end the quartic interaction term of the underlying second quantized Hamilton operator (\ref{Hamiltonian})
is simplified via the prescription \(\hat A \hat B \approx \langle \hat A \rangle \hat B + \langle \hat B \rangle \hat A -\langle \hat A \rangle \langle \hat B \rangle \), where certain choices for the operators \(\hat A\) and \(\hat B\) lead to different variants of mean-field theories. Choosing \(\hat A=\hat \psi_{\uparrow}^\dagger \hat \psi_{\uparrow}\) and \(\hat B = \hat \psi_{\downarrow}^\dagger \hat \psi_{\downarrow}\) or \(\hat A=\hat \psi_{\uparrow}^\dagger \hat \psi_{\downarrow}\) and \(\hat B=\hat \psi_{\downarrow} \hat \psi_{\uparrow}^\dagger\) leads to the Hartree and the Fock channel, respectively, whereas \(\hat A=\hat \psi_{\uparrow}^\dagger \hat \psi_{\downarrow}^\dagger\) and \(\hat B=\hat \psi_{\downarrow} \hat \psi_{\uparrow}\) yields the Bogoliubov channel. With this we obtain the Hartree-Fock-Bogoliubov (HFB) mean-field theory, where the quartic interaction term in (\ref{Hamiltonian}) is approximated according to
\begin{eqnarray}
\label{eq: HFB approximation}
&&	\hat \psi_\uparrow^\dagger
	\hat \psi_\downarrow^\dagger
	\hat \psi_\downarrow
	\hat \psi_\uparrow
	\approx \\
&&
	+
	\left< 	\hat \psi_\uparrow^\dagger	 \hat \psi_\uparrow \right>
	\hat \psi_\downarrow^\dagger \hat \psi_\downarrow
	+
	\hat \psi_\uparrow^\dagger \hat \psi_\uparrow
	\left< 	\hat \psi_\downarrow^\dagger	 \hat \psi_\downarrow \right>
	-
	\left< 	\hat \psi_\uparrow^\dagger	 \hat \psi_\uparrow \right>
	\left< 	\hat \psi_\downarrow^\dagger	 \hat \psi_\downarrow \right> \nonumber \\
&&
    -
	\left< 	\hat \psi_\uparrow^\dagger	 \hat \psi_\downarrow \right>
	\hat \psi_\downarrow^\dagger \hat \psi_\uparrow
	-
	\hat \psi_\uparrow^\dagger \hat \psi_\downarrow
	\left< 	\hat \psi_\downarrow^\dagger \hat \psi_\uparrow \right>
	+
	\left< 	\hat \psi_\uparrow^\dagger	 \hat \psi_\downarrow \right>
	\left< 	\hat \psi_\downarrow^\dagger \hat \psi_\uparrow \right> \nonumber \\
&&
	+
	\left< 	\hat \psi_\uparrow^\dagger \hat \psi_\downarrow^\dagger \right>
	\hat \psi_\downarrow \hat \psi_\uparrow
	+
	\hat \psi_\uparrow^\dagger \hat \psi_\downarrow^\dagger
	\left< 	\hat \psi_\downarrow \hat \psi_\uparrow \right>
	-
	\left< 	\hat \psi_\uparrow^\dagger \hat \psi_\downarrow^\dagger \right>
	\left< 	\hat \psi_\downarrow \hat \psi_\uparrow \right> \, . \nonumber
\end{eqnarray}

In the following we treat the unpolarized case, i.e.~we assume the same number of fermionic atoms for both spin species \(\sigma = \uparrow, \downarrow\). For a homogeneous Fermi gas, where we have $V_{\rm HO}(\vb* x)=0$, the HFB mean-field Hamilton operator reads in Fourier space:
\begin{eqnarray}
\label{HFB Hamiltonian}
    &&\hat H_{\rm HFB}
    =
    \sum_{\vb* k}
    \Bigg[
    \left(\epsilon_k -\mu_{\rm H} \right)
    \hat a_{\vb*k, \uparrow}^\dagger \hat a_{\vb*k, \uparrow}
    -
    \Gamma^{\rm F}
    \hat a_{\vb*k, \uparrow}^\dagger \hat a_{\vb*k, \downarrow}
    \\ &&
    +
    \Delta
	\hat a_{\vb* k,\uparrow}^\dagger
	\hat a_{-\vb* k,\downarrow}^\dagger
    +
    \mbox{h.c.}
    \Bigg]
 %   \nonumber\\ &&
    +
    V
    \Bigg[
    -\frac{(\Gamma^{\rm H})^2}{g}
    +\frac{|\Gamma^{\rm F}|^2}{g}
    -\frac{|\Delta|^2}{g}
    \Bigg] \nonumber
\end{eqnarray}
with the shifted chemical potential $\mu_{\rm H}=\mu - \Gamma^{\rm H}$. Here \(\hat a_{\vb* k,\sigma}^\dagger\) and \(\hat a_{\vb* k,\sigma}\) denote the fermionic creation and annihilation operator with momentum \(\hbar \vb* k\) and spin \(\sigma\), \(\epsilon_k=\hbar^2 \vb* k^2/(2M)\) represents the kinetic energy  of the particles, and \(V\) is the spatial volume. Furthermore, we have introduced the shorthand notation \(\Gamma^{\rm H}=g\langle \hat \psi_\uparrow^\dagger \hat \psi_\uparrow \rangle=g\langle \hat \psi_\downarrow^\dagger \hat \psi_\downarrow \rangle \) and \(\Gamma^{\rm F} = g \langle \hat \psi_\downarrow^\dagger \hat \psi_\uparrow \rangle = g \langle \hat \psi_\uparrow^\dagger \hat \psi_\downarrow \rangle^*\) for the Hartree and the Fock mean-field, whereas \(\Delta=g\langle \hat \psi_\downarrow \hat \psi_\uparrow \rangle=g\langle \hat \psi_\uparrow^\dagger \hat \psi_\downarrow^\dagger \rangle^*\) stands for the Bogoliubov mean-field, which represents the superfluid order parameter.

The diagonalization of the mean-field Hamilton operator (\ref{HFB Hamiltonian}) allows to calculate the grand-canonical free energy \( \mathcal{F}_{\rm HFB} = -\beta^{-1} \ln \mbox{Tr}\big(\,e^{- \beta \hat {H}_{\rm HFB}}\big)\), yielding
\begin{eqnarray}
\label{grand-potential}
    &&\mathcal{F}_{\rm HFB}
    =
    V
    \Bigg[
    -\frac{(\Gamma^{\rm H})^2}{g}
    +\frac{|\Gamma^{\rm F}|^2}{g}
    -\frac{|\Delta|^2}{g}
    \Bigg]
    \\
    &&
    -
    \sum_{\vb* k}
    \Bigg\{
    \frac{E_k^{(+)} +E_k^{(-)}}{2}
    -
    \left(\epsilon_k -\mu_{\rm H} \right)
    \Bigg\}
    \nonumber
    \\
    &&
    -
    \frac{1}{\beta}
    \sum_{\vb* k}
    \Bigg\{
    \ln \left[1+ e^{-\beta E_k^{(+)}} \right]
    +
    \ln \left[1+ e^{-\beta E_k^{(-)}} \right]
    \Bigg\}
    \nonumber
\end{eqnarray}
with the eigenenergies $E_k^{(\pm)} = \left|E_k \pm \left| \Gamma^{\rm F} \right| \right|$, where we have introduced the HB dispersion $E_k= \sqrt{(\epsilon_k -\mu_{\rm H})^2+|\Delta|^2 }$. Subsequently, we consider the yet unknown mean-field energies $\Gamma^{\rm H}, |\Gamma^{\rm F}|$, and $|\Delta|$ as variational parameters and determine them via extremization. At first, we begin with the Fock energy. The extremization condition $\partial \mathcal F_{\rm HFB}/\partial |\Gamma^{\rm F}|=0$ leads to the equation:
\begin{eqnarray}
\label{Fock equation}
    \left|\Gamma^{\rm F}\right|
    =&&
    \frac{g}{2V}
    \sum_{\vb*k}
    \Bigg[
    \frac{1 -\text{sgn}(E_k -\left|\Gamma^{\rm F}\right|)}{2}
    \nonumber \\&&
    -
    \frac{1}{e^{\beta E_k^{(+)}} +1}
    +
    \frac{\text{sgn}(E_k -\left|\Gamma^{\rm F}\right|)}{e^{\beta E_k^{(-)}} +1}
    \Bigg]
    .
\end{eqnarray}
Since the summand at the right-hand side of Eq.~(\ref{Fock equation}) turns out to be positive, we obtain for \(g < 0\) the unique solution \(\Gamma^{\rm F}=0\), i.e.~the Fock energy vasishes.
This result is an immediate consequence of having assumed a contact interaction with vanishing effective range \cite{Czycholl2023}.
Consequently, on the BCS-side, the HFB grand-canonical free energy (\ref{grand-potential}) reduces to the 
corresponding one for the Hartree-Bogoliubov (HB) mean-field theory  \(\mathcal F_{\rm HB}=\mathcal F_{\rm HFB}(\Gamma^{\rm F}=0)\) with \(E_k^{(\pm)}=E_k\). A subsequent extremization of $\mathcal{F}_{\rm HB}$  with respect to the Hartree energy, i.e.~$\partial \mathcal{F}_{\rm HB}/\partial \Gamma^{\rm H} =0$, yields  with the particle density $n = -\partial (\mathcal{F}_{\rm HB}/V)/\partial \mu$
the equation of state
\begin{equation}\label{state-equation}
	n
	=
    \frac{1}{V}
    \sum_{\vb* k}
    \left[
    1
    -
    \frac{\epsilon_k -\mu_{\rm H}}{E_k}
    \tanh\frac{\beta E_k}{2}
    \right]
\end{equation}
and the result $\Gamma^{\rm H} = g n / 2$. Thus, the Hartree energy corresponds to the interaction energy of all Cooper pairs, see e.g.~\cite{Urban,Koehl2015}. Finally, the extremalization with respect to the Bogoliubov energy, i.e.~\(\partial \mathcal{F}_{\rm HB}/\partial |\Delta|=0\), results in an ultraviolet divergent expression. Therefore, the bare interaction strength $g$ needs to be renormalized via \cite{Melo1993}
\begin{eqnarray}
\frac{1}{g} = \frac{1}{V}\sum_{\vb* k} \frac{1}{2\epsilon_k} -\frac{M}{4\pi \hbar^2 a_{\rm s}}\, , 
\end{eqnarray}
where $a_{\rm s}$ denotes the experimentally observable s-wave scatterling length. This leads to the BCS gap equation
\begin{equation}\label{gap-renormalized}
    \frac{M}{2\pi \hbar^2 a_{\rm s}}
    =
    \frac{1}{V}
    \sum_{\vb* k}
    \left(
    \frac{1}{\epsilon_k}
    -
    \frac{1}{E_k}
    \tanh{\frac{\beta E_k}{2}}
    \right)
    \, .
\end{equation}

For fermionic atoms trapped in the harmonic potential \(V_{\rm HO}(\vb* x)=(M/2)\sum_{i=1}^3 (\omega_i x_i)^2\) we apply the local-density approximation (LDA) \cite{Giorini1996}. This amounts to substituting \(\mu \rightarrow \mu(\vb* x) = \mu -V_{\rm HO}(\vb* x)\),
$n \rightarrow  n(\vb* x)$, and $\Delta  \rightarrow \Delta(\vb* x)$, which introduces an additional position dependence. Furthermore, we apply the thermodynamic limit via $\sum_{\vb* k} /V\rightarrow \int d^3k/(2 \pi)^3$. With this we obtain the set of mean-field equations mentioned in the letter.

\end{document}